\documentclass[aps,pre,superscriptaddress,preprintnumbers,amsmath,amssymb,showkeys]{revtex4}
\usepackage{graphicx}% Include figure files
\usepackage{dcolumn}% Align table columns on decimal point
\usepackage{bm}% bold math
\usepackage{float}
\begin{document}
%\section*{High quality GeV proton acceleration driven by an circularly polarized laser pulse}\suppressfloats
%\preprint{APS/123-QED}

\title{High quality GeV proton beams from a density-modulated foil target}
% Force line breaks with \\
%Lines break automatically or can be forced with \\
%\date{\today}% It is always \today, today,
%  but any date may be explicitly specified
\author{T. P. Yu}
\affiliation{Institut f\"{u}r Theoretische Physik I,
Heinrich-Heine-Universit\"{a}t D\"{u}sseldorf, 40225 D\"{u}sseldorf,
Germany} \affiliation{Department of Physics, National University of
Defense Technology, Changsha 410073, China}
\author{M. Chen}
\affiliation{Institut f\"{u}r Theoretische Physik I,
Heinrich-Heine-Universit\"{a}t D\"{u}sseldorf, 40225 D\"{u}sseldorf,
Germany}
\author{A. Pukhov}
\thanks{Address correspondence and reprint requests to: Alexander Pukhov, Institut f\"{u}r Theoretische Physik I,
Heinrich-Heine-Universit\"{a}t D\"{u}sseldorf, 40225 D\"{u}sseldorf,
Germany. E-mail: pukhov@tp1.uni-duesseldorf.de}
\affiliation{Institut f\"{u}r Theoretische Physik I,
Heinrich-Heine-Universit\"{a}t D\"{u}sseldorf, 40225 D\"{u}sseldorf,
Germany}
%/
%/\author{F. Q. Shao}
%/\affiliation{Department of Physics, National University of Defense
%/Technology, Changsha 410073, China}

\begin{abstract}
We study  proton acceleration from a foil target with a transversely
varying density using multi-dimensional Particle-in-Cell (PIC)
simulations. In order to reduce electron heating and deformation of
the target, circularly polarized Gaussian laser pulses at
intensities of the order of $10^{22} \mathrm {Wcm}^{-2}$ are used.
It is shown that when the target density distribution fits that of
the laser intensity profile, protons accelerated from the center
part of the target have quasi-monoenergetic spectra and are well
collimated. In our two-dimensional PIC simulations, the final peak
energy can be up to 1.4 GeV with the full-width of half maximum
divergence cone of less than $4^\circ$. We observe highly efficient
energy conversion from the laser to the protons in the simulations.
\end{abstract}
%\pacs{52.40Nk, 52.35.Mw, 52.57.Jm, 52.65.Rr}
% PACS, the Physics and Astronomy Classification Scheme. %%
%Use showkeys class option if keyword
%display desired
\maketitle

\section{\label{sec:level1}INTRODUCTION\protect}

Plasma-based particle acceleration has recently demonstrated an
impressive progress. Monoenergetic electron beams with up to GeV
energy have already been observed in recent experiments (Malka
{\itshape et al}., 2002; Leemans {\itshape et al}., 2006). Energetic
ion bunches, especially monoenergetic proton beams, have also been
obtained in some laboratories (Schwoerer {\itshape et al}., 2006;
Fuchs {\itshape et al}., 2006). It is believed that Target Normal
Sheath Acceleration (TNSA) (Wilks {\itshape et al}, 2001) is a
dominant proton acceleration mechanism when the laser intensity is
below some $10^{20} \mathrm {Wcm}^{-2}$ level. However, up to now
the energy of ion beams in the "TNSA" regime is only about a few
tens MeV with a low energy conversion efficiency ($\leq 1\%$), which
is insufficient for most of the envisioned practical application,
such as, e.g. the tumor therapy (Bulanov {\itshape et al}., 2002)
and proton imaging (Borghesi {\itshape et al}., 2003). In the past
few years, numerous experimental and theoretical studies (Pegoraro
{\itshape et al}., 2004; Yin {\itshape et al}., 2006; Nickles
{\itshape et al}., 2007; Flippo {\itshape et al}., 2007; Ma
{\itshape et al}., 2009; Yu {\itshape et al}., 2009) have been
devoted to improving the beam quality. Recently, a new ion
acceleration mechanism, Radiation Pressure Acceleration (RPA)
(Robinson {\itshape et al}., 2008) has attracted a lot of attention
due to its potential to directly transfer the momentum of the laser
light to the thin target as a whole. A complete switch from the TNSA
to the RPA regime occurs at a laser intensity of $10^{21} \mathrm
{Wcm}^{-2}$ (Robinson {\itshape et al}., 2008) for a circularly
polarized (CP) laser pulse, which opens a new roadmap to high
quality ion acceleration.

In the "TNSA" regime, a linearly polarized (LP) laser pulse is
usually employed due to its superiority in producing hot electrons.
Instead, a CP laser pulse with a high peak intensity is more
efficient in the generation of quasi-monoenergetic proton beams
(Zhang {\itshape et al}., 2007; Yan {\itshape et al}., 2008). It is
because of the absence of the oscillating component in the
ponderomotive force described below:
\begin{eqnarray}
f_p^L&=&{e^2\over{4\gamma m\omega^2}}{\partial\over\partial z}E^2(z)(1+\cos(2\omega t))\nonumber\\
f_p^C&=&{e^2\over{4\gamma m\omega^2}}{\partial\over\partial
z}E^2(z),
\end{eqnarray}

\noindent where $m$, $e$ and $\omega$ are electron static mass,
electron charge and laser frequency, respectively. $\gamma$ is the
relativistic factor and $E(z)$ is the laser electric field
component. The oscillating part in the ponderomotive force of the LP
laser pulse, $f_p^L$, can excite a strong oscillation of the
electrons. As a result, much more hot electrons are produced, which
are essential for the TNSA mechanism. However, for a CP laser pulse,
the ponderomotive force has no such a term, but only the time
average or zero-frequency component. The strong force directly
pushes the electrons inwards the target and forms strong electric
fields behind the laser front (Chen {\itshape et al}., 2008).
Choosing the appropriate laser and target parameters, one can expect
that quasi-monoenergetic proton beams can be produced by these
fields (Chen {\itshape et al}., 2009).

To describe the CP laser interaction with an ultra-thin foil, we
assume that the force applies to the whole target and that the foil
still stays intact during the full laser interaction time. As a
result, the target is pushed forward as a whole. Using a simple 1D
analytical model, we obtain for the target velocity the following
equation (Robinson {\itshape et al}., 2008):
\begin{eqnarray}
{dv\over dt}={1\over 2\pi m_in_ic}{E^2(t,x,r)\over l_0}{1\over
\gamma^3}{(1-v)\over (1+v)},
\end{eqnarray}

\noindent where $m_i$, $n_i$ and $l_0$ are ion mass, plasma density
and target thickness, respectively. $v$ is the target velocity
normalized by light speed $c$ and $E(t,x,r)$ is the laser electric
field component. From the formula, we can see that the acceleration
structure is dependent on two factors: target parameters
($m_i,n_i,l_0$) and laser transverse profile ($E$). For a usual
uniform density flat target (UFT) irradiated by a Gaussian laser
pulse, the acceleration structure will be soon destroyed due to the
target deformation. It is because different parts of the target has
experienced different acceleration forces. In order to avoid the
target deformation, Chen {\itshape et al}. proposed a shaped foil
target (SFT) with an transversely varying thickness (Chen {\itshape
et al}., 2009). PIC simulations show that the scheme can
 suppress both the target deformation and heating efficiently.

In this paper, we suggest an alternative method  to produce the high
quality proton beams. In our case, the initial foil target is a flat
one, but the transverse plasma density follows a Gaussian
distribution to match the laser intensity profile. A CP laser pulse
is employed and is normally incident on this density-modulated foil
target (DMFT). 2D and 3D simulations have been performed, which show
that protons from the center part of the target can be accelerated
monoenergetically and are well collimated in the forward direction.
In our simulations, we observe the final peak energy as high as 1.4
GeV with the full-width of half maximum divergence cone of less than
$4^\circ$.

\section{PIC SIMULATIONS AND DISCUSSIONS}

We first present a 2D simulation of the scenario using the fully
electromagnetic PIC code VLPL (Virtual Laser Plasma Laboratory)
(Pukhov, 1999). The simulation box is $48 \lambda$ long and $32
\lambda$ wide ($\lambda=1.0\mu m$ is the wavelength), which consists
of $4800 \times 320$ cells, and contains more than $4.2\times10^6$
macroparticles. The foil target is initially located between $x=5.0
\lambda$ and $5.3 \lambda$. A CP laser pulse with a Gaussian profile
in space and a trapezoidal profile (linear growth - plateau - linear
decrease) in time is normally incident on the foil target:
 %\makeatletter
%\let\@@@alph\@alph
%\def\@alph#1{\ifcase#1\or \or $'$\or $''$\fi}\makeatother
%\begin{subnumcases}
%{a=} a_0\exp(-{y^2\over \sigma_l^2})t, \quad 0\leq t<1T\\
%a_0\exp(-{y^2\over \sigma_l^2}),\quad 1T\leq t\leq9T,\\
%a_0\exp(-{y^2\over \sigma_l^2})(10-t),\quad 9T<t\leq10T,
%\end{subnumcases}
%\makeatletter\let\@alph\@@@alph\makeatother

\begin{equation} \label{eq:4}
a=\left\{ \begin{aligned}
         & a_0\exp(-\frac{y^2}{\sigma_L^2})t, \quad 0\leq t<1T\\
         & a_0\exp(-\frac{y^2}{\sigma_L^2}),\quad 1T\leq t\leq6T,\\
         & a_0\exp(-\frac{y^2}{\sigma_L^2})(8-t),\quad 6T<t\leq7T,
         \end{aligned} \right.
\end{equation}

\noindent where $a_0=100$ is the  laser intensity normalized by
$Ec/m\gamma c$, $\sigma_l=8 \lambda$ is the focal spot radius,
$T=3.3fs$ is the laser cycle. The initial plasma density follows a
transverse Gaussian distribution to match the laser intensity
profile, as shown in Fig.1. The profile of the modulated density is
defined by $\sigma_d=7\lambda$. The maximal density is $100n_c$
while the cutoff is $20n_c$, where $n_c$ is the critical density.
The transverse boundary conditions are periodic, while both the
front and back boundaries absorb outgoing radiation and particles
(Pukhov, 2001). Considering the plasma expansion into vacuum, we
provide an appropriate vacuum gap (longer than 42$\mu m$) between
the target and the right boundary.

Fig.~2 (a) shows the proton energy spectra at $t=10~T,~~ 20~T,~~
30~T$ and $40~T$.  Here, the leading edge of the laser pulse reaches
the target at about $t=5~T$. A clear quasi-monoenergetic peak can be
seen in each spectrum. At an early time, $t=10~T$, the peak energy
is about 200~MeV with a very narrow energy spread. As time goes on,
the proton energy increases. At the time $t=40~T$, the peak is still
very clear although the spectrum shows a relatively wide energy
spread. By this time, the peak energy is up to 1.2GeV and contains
$6.5\times10^7$ protons while the cutoff energy is about 1.5GeV. The
number of the protons within the energy range 0.8~GeV$-$1.3~GeV  is
$2.0\times10^{10}$. The monoenergetic peak is accelerated up to
1.4~GeV with the full-width of half maximum divergence cone of less
than $4^\circ$ at the time $t=50~T$ (165fs).

The proton energy as a function of the divergency angle is shown in
Fig.~3.  It is easy to see from both of the frames that there exists
a bunch of protons with a relatively high energy and a low
divergency. At $t=25~T$, the clump is composed of  protons within
the energy range 0.65~GeV$-$0.85GeV. However, at a later time
$t=40~T$, the same protons
 are shifted to the energy range 0.8~GeV$-$1.3~GeV. The average divergency angle for all
these high quality protons is about $2.2^\circ$ at $t=25~T$ and
$3.5^\circ$ at $t=40~T$. Here, the average divergency is calculated
as following:
\begin{eqnarray}
\theta_{ave}&=&\sqrt[]{\sum\limits_{i=1,..N}(\theta_i)^2/N},
\quad\theta_i=tan^{-1}(p_y/p_x),
\end{eqnarray}

\noindent where $N$ is the total numbers of the high quality protons,
$p_x$ and $p_y$ are the momentum component in $X-$ and $Y-$direction,
respectively.

Fig.~4 presents snapshots of the laser intensity and proton
acceleration at $t=25~T$ and $t=40~T$. Because of the lower density
at the target wing, the ultra-intense laser pulse can easily
penetrate it and then propagate into the vacuum behind the target.
On the contrary, the center part of the target in the range between
$y=10\lambda$ to $22\lambda$ is directly pushed forward by the
strong ponderomotive force $f_p^C$. As a result, the laser intensity
shows a clear inverted cone distribution, as shown in Fig.~4(a) and
4(b). It is this inverse cone that keeps the clump together.
According to the formula (2), the protons from the center part will
experience a uniform acceleration so that a good acceleration
structure survives for a long time, as plotted in Fig. 4(c) and
4(d). Our simulation results qualitatively agree with the above 1D
analytical model. Additionally, we also record the proton energy
distribution in the space, see Fig.~4(e) and 4(f). By comparing the
density distributions, one can easily observe the high quality
proton clump mentioned above. The "radius" of the clump is about
$6\lambda$ at $t=25~T$ and $8\lambda$ at $t=40~T$, which is
approximately equal to the laser focus.

3D PIC simulations have also been performed to check the proton
acceleration. Here, both of the shape of the DMFT and the laser
profile are the same except the initial target position and
$\sigma_d$. In the 2D case, the target is located at $x=5\lambda$
with $\sigma_d = 7\lambda$ while in the 3D case they are $2\lambda$
and $6\lambda$, respectively. The pulse duration in the 3D
simulations is $7T$, which corresponds to a trapezoidal profile
1T--5T--1T. To reduce the computational time, the full simulation
box has a size $X\times Y\times
Z=25\lambda\times27\lambda\times27\lambda$ sampled by a grid of
$2500\times225\times225$ cells. Fig. 2(b) shows the proton energy
spectra at $t=10~T$, $15~T$ and $20~T$. An obvious energy peak can
be observed there. At t=20T, the spectrum shows a peak with the
energy of 0.9~GeV corresponding to $5.4\times10^9$ protons. The
total number of the protons with an energy larger than 0.6 GeV is
about $1.1\times10^{12}$, which contains a total energy of $155J$.
The energy conversion efficiency from the laser pulse to these
protons is up to 27.1\%, which is much higher than that obtained in
most other mechanism regimes.

Fig. 5 presents the spatial density distribution of the protons. We
can see that the target can keep a good acceleration structure. The
simulations confirm the results in the above 2D simulations.
Additionally, we also observe the expected proton clump behind the
target in the 3D simulations, as shown in Fig. 5(b)--(e). The radius
of the clump is about $4.5\lambda$, which is smaller than the laser
focus. It may be due to the easier dispersion of the protons in the
3D condition. In fact, the size of the clump depends on the cutoff
density, laser focus as well as $\sigma_d$. When $\sigma_d$ is
matched with the laser focus, for a lower cutoff density more
protons from the wing target will be uniformly accelerated, which
leads to a wider clump radius. On the contrary, these wing protons
experienced inhomogeneous forces and would be filtered by the laser
pulse. As a preliminary estimation, the optimal cutoff density is
half of the maximum, that is $50n_c$ in our case.

\section{COMPARISON OF THE BEAM QUALITY WITH OTHER TARGET PROFILES}

We compare our target with some other profiles, as shown in Fig.
6(a). Among them, the case 2 is just the usual flat foil target
(UFT) with the density of $100n_c$, while case 3 is another
specially-organized foil target with a density of the transverse
linear distribution. Both of the maximal density and cutoff density
in the cases 1 and 3 are the same. Case 4 is the SFT presented by
Chen {\itshape et al.} (Chen {\itshape et al}., 2009), where the
foil thickness is matched to the laser intensity profile. For the
convenience of comparison, here the SFT is made with a matched
profile (corresponding to a cutoff thickness of $0.06\lambda$) so
that the whole target contains the same number of protons as that in
our case. All these targets are located at the same position with
the same thickness (for the SFT, it is the maximal thickness) and
are irradiated by the same CP laser pulses. In order to save the
computational time, we only perform 2D PIC simulations.

Fig. 6(b) presents the spectra of all the protons from the target at
$t=25~T$. Obviously, only the spectra in the cases 1 and 4 show a
quasi-monoenergetic peak structure. That is because the both targets
employ a Gaussian profile to match the laser profile, which leads to
the uniform acceleration of the target as a whole. In the UFT case,
the acceleration structure is destroyed very soon and the spectrum
shows an exponential decay. In the case 3 we do observe formation of
an inverse cone in the laser intensity behind the target. Yet,
different parts of the target experience different acceleration,
because the target profile is not matched with that of the laser.
Due to the transverse linear distribution of the density, the energy
spectrum is not an exponential one, but rather shows a nearly flat
distribution. When we compare the DMFT case (the case 1) with the
SFT case (the case 4), we mention that there is almost no difference
for the distribution of the high energy protons except that, in our
case, the number of low energy protons is reduced and more energy is
focused on the clump mentioned above. Finally, the energy conversion
efficiency from the laser pulse to the high quality protons is
highly enhanced.

Finally, we compare the divergence angle for these cases, as shown
in Fig.~6(c). As expected, both our DMFT case and the SFT can
produce a proton beam with a better collimation. On the contrary,
the angle distribution for the UFT shows a larger divergency. That
is because the electrons in the UFT are easily scattered by the
laser and spread into the vacuum.  However, in the DMFT case and SFT
case, due to the uniform acceleration, all parts of the target are
pushed forward as a whole. Then, the protons have a low divergency
angle. On the other hand, compared with the SFT, the proton
collimation in the DMFT case is much better.  The number of protons
with the full-width of half maximum divergence cone of less than
$2.7^\circ$ in the SFT is about $1.8\times10^{10}$, which is only
about 80\% of that in the DMFT case. This should be attributed to
the inverse cone of laser intensity formed behind the DMFT, which
keeps the protons together. On the whole, the beam quality in our
case is
 higher than that in the SFT and much better than that in the UFT.

\section{CONCLUSIONS}
In conclusion, we study proton acceleration from a density-modulated
foil target. In order to avoid the deformation of the target, the
density follows a transverse Gaussian distribution to match the
laser intensity profile. Meanwhile, a CP laser pulse at intensities
of $2.72\times10^{22} \mathrm {Wcm}^{-2}$ is employed to push the
target uniformly. Our 2D and 3D simulations demonstrate generation
of the high quality proton beams. A proton clump with a higher
energy and better collimation is observed behind the target, whose
radius is about equal to that of the laser focus in the 2D
simulations. The peak energy of the quasi-monoenergetic protons can
be up 1.4~GeV. The corresponding full-width of half maximum
divergence cone is less than $4.0^\circ$. The energy conversion
efficiency can be up to 27.1\% in the 3D simulation. By comparison
with some other reference targets, such as the UFT and the SFT, both
the acceleration structure and the beam quality as well as the
energy conversion efficiency in the DMFT case are further improved.

\begin{acknowledgments}\suppressfloats
% Note for American journals do not put the title of people anywhere in the manuscript.
We thank Prof. F.Q.Shao and Dr. Y.Y.MA for their helpful discussions
on this subject. This work is supported by the DFG programs GRK1203
and TR18. T.P.Yu thanks the scholarship awarded by China Scholarship
Council (CSC NO. 2008611025). M. Chen acknowledges the support by
the Alexander von Humboldt Foundation.
\end{acknowledgments}

\newpage
\section*{Figure Captions}\suppressfloats
Fig.\ 1. Schematic diagram of the DMFT case. The curved line shows
the density distribution along the transverse direction. The dashed
line indicates the cutoff density of the target. $\sigma_d$ defines
the transverse density profile. A CP laser pulse
is incident on the foil target from the left boundary.\\

Fig.\ 2. Proton energy spectra for the DMFT case in the 2D
simulations (a) and 3D simulations (b). Here, both of the shape of
the DMFT and the laser profile are the same except the initial target position and $\sigma_d$.\\

Fig.\ 3. Proton energy as a function of the divergency angle for the
DMFT in the 2D simulation at (a) $t=25~T$  and (b)
$t=40~T$.\\

Fig.\ 4. Spatial distributions of the laser intensity
$(E_y^2+E_z^2)$ for the DMFT case in the 2D simulation at (a) t=25T
and (b) t=40T . Spatial
 density distributions of protons for the DMFT case in the 2D simulation at (c) t=25T and (d) t=40T.
Spatial energy distributions of protons for the DMFT case in the 2D
simulation at (e) t=25T and (f) t=40T. The uniformly accelerated
protons with up to GeV energy are observed.\\

Fig.\ 5. Spatial density distributions of protons for the DMFT case
in the 3D simulation at $t=5~T$, $10~T$, $15~T$ and $20~T$. A clear
proton clump formed behind the target can be easily
distinguished from (b), (c) and (e).\\

Fig.\ 6. Comparison among different target profiles (a). Here all
the laser parameters are the same. For the cases 1 and 3, the
density follows a transverse Gaussian distribution and linear
distribution, respectively. Both of the maximal density and cutoff
density are the same. For the case 2, it is a usual flat target with
a uniform density of $100n_c$. For the case 4, the target thickness
follows the transverse Gaussian distribution while the density is
also uniform ($100n_c$). Proton energy spectra (b) and divergency
angle (c) at $t=25~T$ are shown in the second and third frames.

\newpage
\begin{figure}\suppressfloats
\includegraphics[width=10cm]{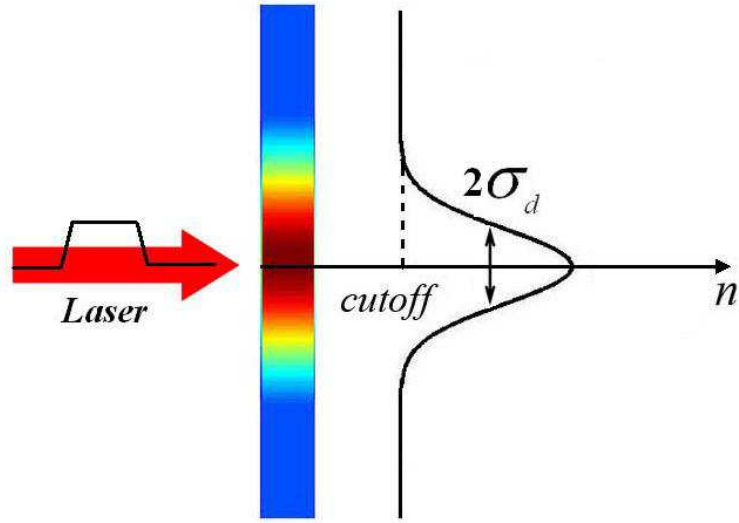}%
\caption{\label{f1} Schematic diagram of the DMFT case. The curved
line shows the density distribution along the transverse direction.
The dashed line indicates the cutoff density of the target.
$\sigma_d$ defines the transverse density profile. A CP laser pulse
is incident on the foil target from the left boundary.}
\end{figure}

\newpage
\begin{figure}\suppressfloats
\includegraphics[width=16cm]{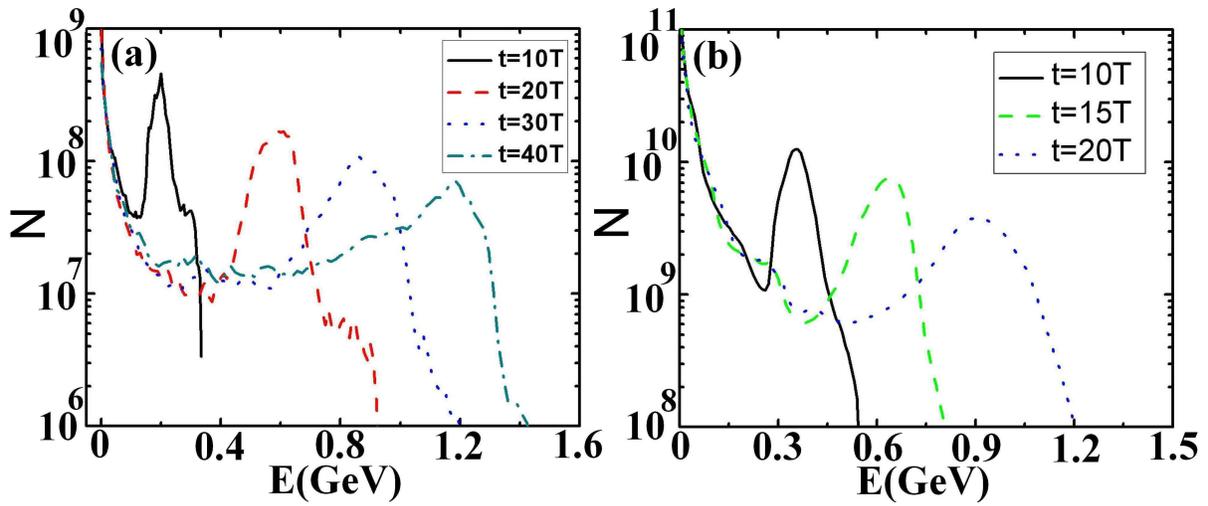}%
\caption{\label{f2} Proton energy spectra for the DMFT case in the
2D simulations (a) and 3D simulations (b). Here, both of the shape
of the DMFT and the laser intensity as well as the pulse duration
are the same except the initial target position and $\sigma_d$.}
\end{figure}

\newpage
\begin{figure}\suppressfloats
\includegraphics[width=16cm]{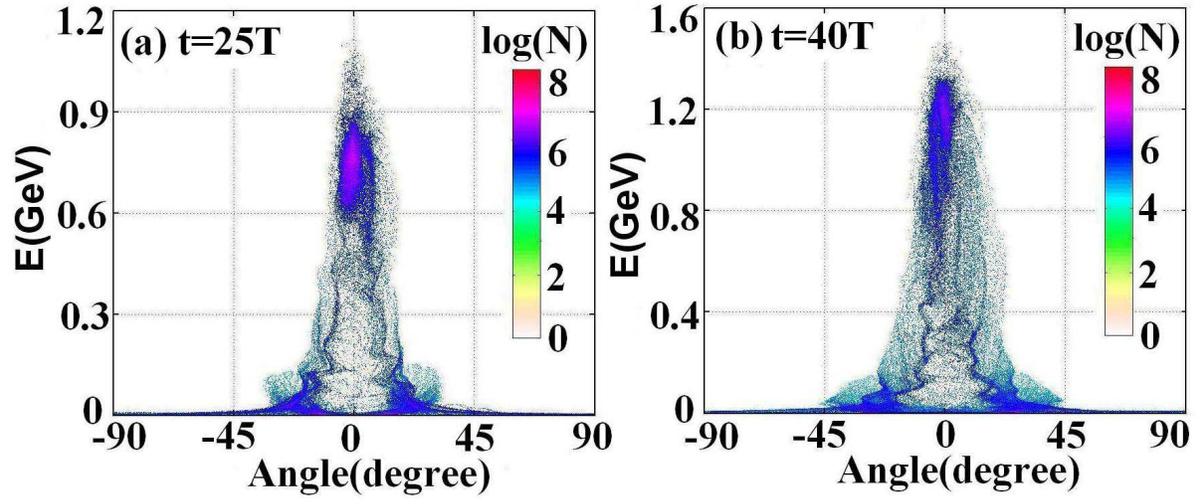}%
\caption{\label{f3} Proton energy as a function of the divergency
angle for the DMFT in the 2D simulation at (a) $t=25~T$  and (b)
$t=40~T$.}
\end{figure}

\newpage
\begin{figure}\suppressfloats
\includegraphics[width=16cm]{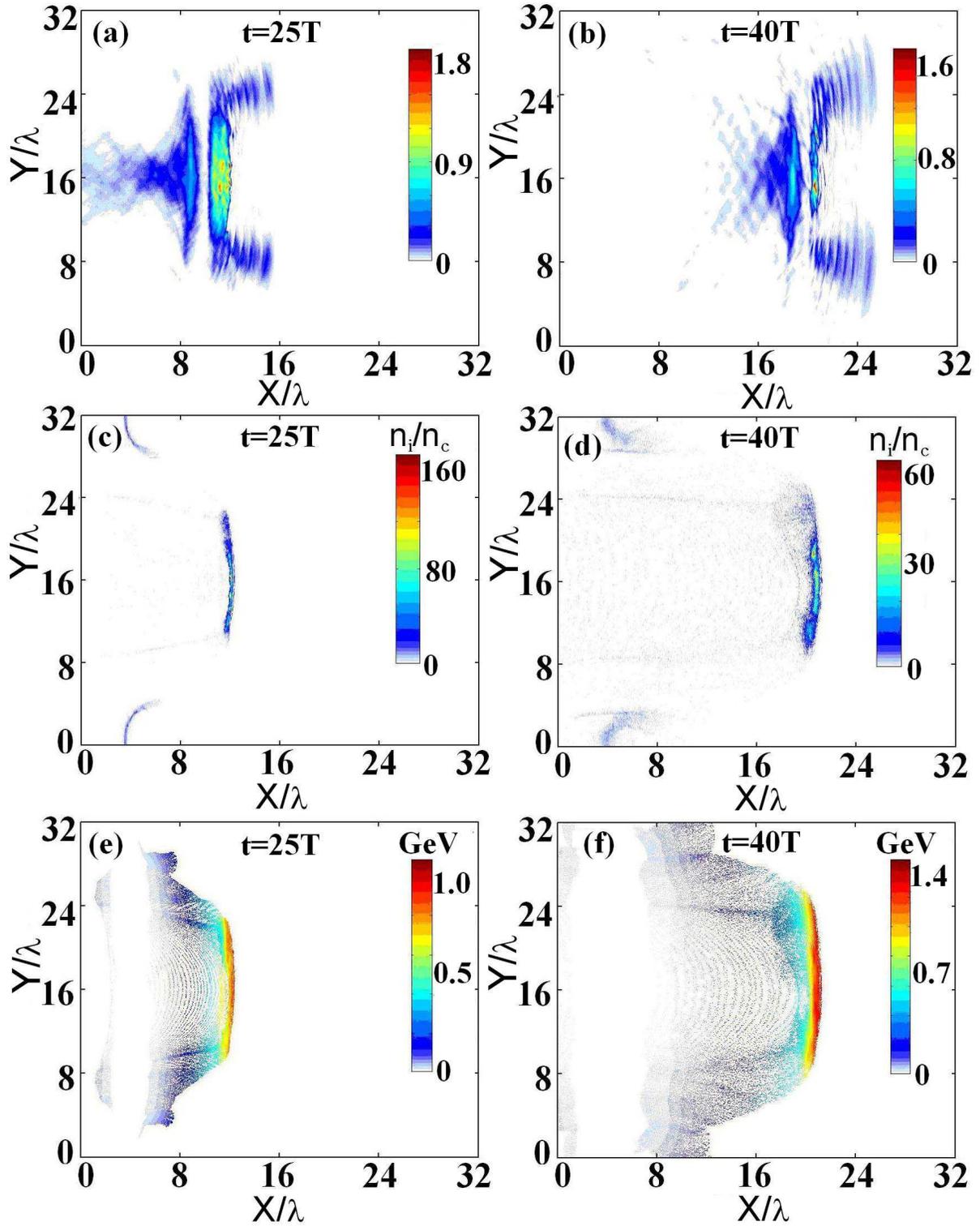}%
\caption{\label{f4} Spatial distributions of the laser intensity
$(E_y^2+E_z^2)$ for the DMFT case in the 2D simulation at (a) t=25T
and (b) t=40T . Spatial
 density distributions of protons for the DMFT case in the 2D simulation at (c) t=25T and (d) t=40T.
Spatial energy distributions of protons for the DMFT case in the 2D
simulation at (e) t=25T and (f) t=40T. The uniformly accelerated
protons with up to GeV energy are observed.}
\end{figure}

\newpage
\begin{figure}\suppressfloats
\includegraphics[width=16cm]{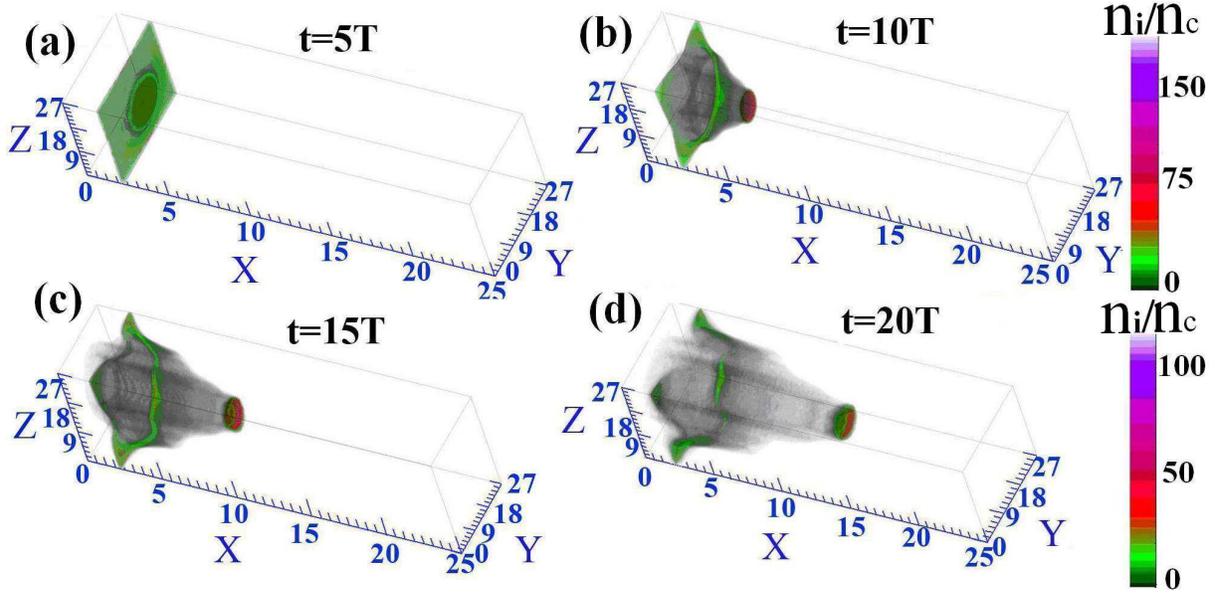}%
\caption{\label{f5} Spatial density distributions of protons for the
DMFT case in the 3D simulation at $t=5~T$, $10~T$, $15~T$ and
$20~T$. A clear proton clump formed behind the target can be easily
distinguished from (b), (c) and (e).}
\end{figure}

\newpage
\begin{figure*}[!htb]\suppressfloats
\includegraphics[width=16cm]{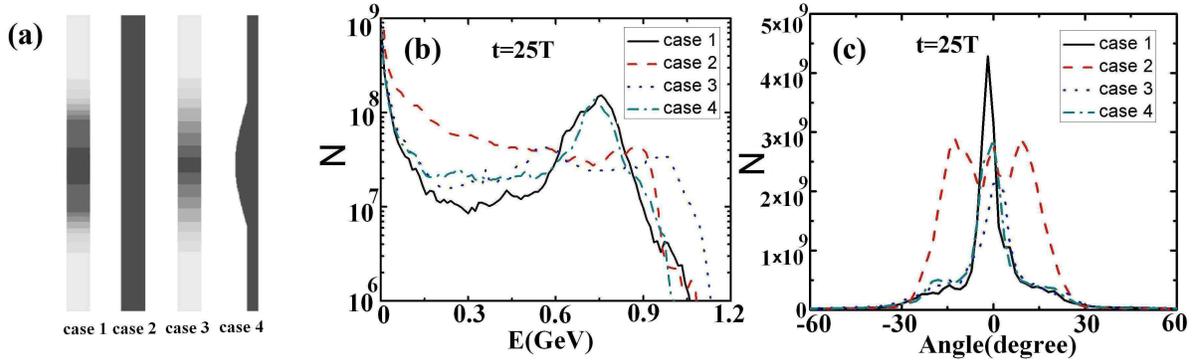}%
\caption{\label{f6} \texttt{}Comparison among different target
profiles (a). Here all the laser parameters are the same. For the
cases 1 and 3, the density follows a transverse Gaussian
distribution and linear distribution, respectively. Both of the
maximal density and cutoff density are the same. For the case 2, it
is a usual flat target with a uniform density of $100n_c$. For the
case 4, the target thickness follows the transverse Gaussian
distribution while the density is also uniform ($100n_c$). Proton
energy spectra (b) and divergency angle (c) at $t=25~T$ are shown in
the second and third frames.}
\end{figure*}
\end{document}